# Stabilization of a honeycomb lattice of IrO$_6$ octahedra in superlattices with ilmenite-type MnTiO$_3$

Kei Miura[1], Kohei Fujiwara[1*], Kei Nakayama[2], Ryo Ishikawa[2,3], Naoya Shibata[2] and Atsushi Tsukazaki[1,4]

[1]*Institute for Materials Research, Tohoku University, Sendai 980-8577, Japan*

[2]*Institute of Engineering Innovation, The University of Tokyo, Tokyo 113-8656, Japan*

[3]*PRESTO, Japan Science and Technology Agency, Kawaguchi 332-0012, Japan*

[4]*Center for Spintronics Research Network (CSRN), Tohoku University, Sendai 980-8577, Japan*

[*]Author to whom correspondence should be addressed: kfujiwara@imr.tohoku.ac.jp

**Abstract**

In the quest for quantum spin liquids, thin films are expected to open the way for the control of intricate magnetic interactions in actual materials by exploiting epitaxial strain and two-dimensionality. However, materials compatible with conventional thin-film growth methods have largely remained undeveloped. As a promising candidate towards the materialization of quantum spin liquids in thin films, we here present a robust ilmenite-type oxide with a honeycomb lattice of edge-sharing $IrO_6$ octahedra artificially stabilized by superlattice formation with an ilmenite-type antiferromagnetic oxide $MnTiO_3$. The stabilized sub-unit-cell-thick Mn–Ir–O layer is isostructural to $MnTiO_3$, having the atomic arrangement corresponding to ilmenite-type $MnIrO_3$ not discovered yet. By spin Hall magnetoresistance measurements, we found that antiferromagnetic ordering in the ilmenite Mn sublattice is suppressed by modified magnetic interactions in the $MnO_6$ planes via the $IrO_6$ planes. These findings lay the foundation for the creation of two-dimensional Kitaev candidate materials, accelerating the discovery of exotic physics and applications specific to quantum spin liquids.

**Introduction**

Recent progress in the physics of quantum spin liquids has intensified search for new candidate materials[1]. In the materials design, an exactly solvable $S = 1/2$ spin model on a two-dimensional honeycomb lattice has given a rigorous theoretical framework, which was firstly proposed by Kitaev[2] and later reformalized with realistic materials by Jackeli and Khaliullin[3]. The key ingredient of the model is a honeycomb lattice of edge-sharing metal-anion octahedra comprising of $Ru^{3+}$ or $Ir^{4+}$ ions with a $d^5$ electron configuration, which produces bond-specific Ising-like interactions, called Kitaev-type interactions, giving rise to strong quantum fluctuations of (pseudo)spins on the honeycomb lattice[1,3–5]. In $\alpha$-RuCl$_3$ (refs. [6,7]) and iridates like Na$_2$IrO$_3$ (refs. [8,9]), $\alpha$-, $\beta$-, and $\gamma$-Li$_2$IrO$_3$ (refs. [9–13]), and H$_3$LiIr$_2$O$_6$ (ref. [14]) with such a geometry, possibilities of quantum spin liquids via Kitaev-type interactions have been argued intensively. In tandem with theoretical approaches[3–5], advanced experiments on bulk crystals, such as magnetic resonant x-ray diffraction (XRD)[10,13], thermal Hall effect[15], and nuclear magnetic resonance[14], have contributed to clarifying complex magnetic phases, as well as identifying the quantum spin liquid state.

One of the major challenges in this research field is to incorporate those materials into thin films, especially atomically thin monolayer forms. Thin-film techniques are not only essential for eventual practical applications but also effective for providing additional degrees of freedom for controlling magnetic interactions in actual materials. For example, epitaxial strain at the interface and dimensionality control[16–18] can offer new options to address the issue that the expected Kitaev-type interactions are often broken by the antiferromagnetic ordering due to structural distortion and interlayer coupling[6–13,19,20]. However, it is quite difficult to prepare high-quality films of $\alpha$-RuCl$_3$ (refs. [6,7]) and iridates with H and Li (refs. [10–14]) that have so far gained attention as the bulk candidate materials. This is because their volatile and/or diffusible nature must entail extremely careful control of stoichiometry and suppression of inter-diffusion for reproducing the bulk properties. Considering these problems in known Kitaev candidate materials, we aimed to find a new crystal system applicable for the thin-film research and focused on ilmenite-type $AB$O$_3$ compounds (where $A$ and $B$ are metal cations) containing Ir.[19,20] Recently, Haraguchi *et al.* reported the bulk synthesis of ilmenite-type ZnIrO$_3$ and MgIrO$_3$, which possess a honeycomb lattice of edge-sharing IrO$_6$ octahedra in *ab* plane[19] (see Fig. 1(a) for the ilmenite-type structure). The authors discussed an XY-like magnetic anisotropy and a tilting magnetic structure that is possibly related to Kitaev-type interactions[19]. In this study, starting from ilmenite-type MnTiO$_3$ for which the single-crystalline film growth on Al$_2$O$_3$(0001) has been established[21,22], we attempted to stabilize Ir at the *B*-site of ilmenite-type Mn$B$O$_3$, as shown in Fig. 1(a). By taking advantage of epitaxial strain in superlattices with ilmenite-type MnTiO$_3$, which shares the *A*-site MnO$_6$ plane, we materialized the honeycomb lattice of edge-sharing IrO$_6$ octahedra in the ilmenite lattice. This ultrathin-film form of iridate will serve as an intriguing two-dimensional platform for pursuing the physics of Kitaev materials.

**Results**

**Structural characterization by x-ray diffraction (XRD).** The films were grown on Al$_2$O$_3$(0001) substrates by pulsed-laser deposition using a KrF excimer laser (see Methods). Our early attempt to grow ilmenite-type MnIrO$_3$ directly on Al$_2$O$_3$(0001) was not successful, resulting in phase-separated films of Mn$_3$O$_4$ and IrO$_2$ (see Fig. S1 in Supporting Information for the XRD pattern). We then took a superlattice approach[16–18] based on ilmenite-type MnTiO$_3$ thin films[21,22]. In the superlattice structure, the common $A$-site MnO$_6$ plane is expected to promote the ilmenite-type stacking of MnO$_6$ and IrO$_6$ planes along the $c$-axis direction (the left of Fig. 1(a)). A schematic structure of a typical superlattice is given in Fig. 1(b), where a MnTiO$_3$ buffer layer with a thickness between 4.3 nm and 7.8 nm is initially grown on Al$_2$O$_3$(0001), followed by alternate deposition of much thinner Mn−Ir−O and MnTiO$_3$ layers. The film thickness of 1.4 nm is roughly equal to typical $c$-axis parameters of ilmenite-type oxides. In a 0.5-nm-thick Mn−Ir−O (cap)/[4.3-nm-thick MnTiO$_3$/0.5-nm-thick Mn−Ir−O]$_{15}$/5.7-nm-thick MnTiO$_3$(buffer) film (superlattice cycle number $n$ = 15 in Fig. 1(b)), the overall chemical composition of the film as evaluated by energy-dispersive x-ray spectroscopy (EDX) was Mn : Ti : Ir = 50.4 : 45.1 : 4.5, in good agreement with the ideal composition of Mn : Ti : Ir = 50.0 : 44.9 : 5.1 expected from the thickness ratio designed. In the out-of-plane XRD pattern shown in Fig. 1(c), four main diffraction peaks are observed from the film, which are assignable to (000$\underline{3n}$) (*n*: natural number) of the ilmenite-type structure. The appearance of satellite peaks near (0006) and (00$\underline{12}$) peaks (indicated by red arrows) reflects periodic modulation in the x-ray scattering cross-section by the superstructure of MnTiO$_3$ and Mn−Ir−O layers, evidencing that an alloy compound of Mn(Ti,Ir)O$_3$ is not formed. Assuming the abrupt MnTiO$_3$/Mn−Ir−O interface without inter-diffusion, the average in-plane and out-of-plane lattice parameters of the single unit of the superlattice, i.e., [4.3-nm-thick MnTiO$_3$/0.5-nm-thick Mn−Ir−O], are found to be *a* = 5.13 Å and *c* = 14.29 Å from the reciprocal space mapping around (02$\bar{2}$10) (Fig. 1(d)). Since these values differ very little from *a* = 5.1396 Å and *c* = 14.29 Å of MnTiO$_3$ (JCPDS PDF No. 00-029-0902), the lattice parameters of Mn−Ir−O are comparable to those of MnTiO$_3$. Also, the three-fold symmetry of the (10$\bar{1}$4) diffraction peak is detected in the in-plane azimuthal $\phi$ scan measurements (Fig. 1(e)), confirming the epitaxial orientation relationship of the film [10$\bar{1}$0](0001) // Al$_2$O$_3$ [10$\bar{1}$0](0001). Combined together, these results indicate that the film grows as a *c*-axis oriented, single-crystalline ilmenite-type oxide with a periodically modulated internal structure.

We examined the critical $d_{\text{MIO}}$ value for maintaining the single-crystalline film growth by varying the thickness of Mn−Ir−O layers ($d_{\text{MIO}}$) in the [4.3-nm-thick MnTiO$_3$/$d_{\text{MIO}}$-nm-thick Mn−Ir−O]$_{15}$ superlattice. As shown in Figs. 2(a) and 2(b), the films with $d_{\text{MIO}}$ = 0.5 nm (identical to the sample in Fig. 1) and 1.0 nm exhibit clear (0006) peaks associated with satellite peaks (red arrows). In particular for $d_{\text{MIO}}$ = 0.5 nm, thickness fringes appear around the (0006) peak, which indicate that the total film thickness is uniform over the entire film. Given that the observed satellite peaks are superlattice reflections, the single-unit lengths of the superlattices are calculated to be 4.5 nm for $d_{MIO}$ = 0.5 nm and 5.6 nm for $d_{MIO}$ = 1.0 nm, which coincide well with the designed values of 4.8 nm (= 4.3 + 0.5 nm) and 5.3 nm (= 4.3 + 1.0 nm), respectively. In the thickest film with $d_{MIO}$ = 1.4 nm (Fig. 2(c)), however, superlattice reflections become indiscernible along with the occurrence of a diffraction peak of segregated $IrO_2$ impurities; the designed superlattice structure is no longer formed for $d_{MIO}$ = 1.4 nm. The concomitant decrease in the (0006) diffraction intensity implies that the basal $MnTiO_3$ layers in the superlattice are also rather disordered. Therefore, the upper bound of $d_{MIO}$ for stabilizing $IrO_6$ planes in the ilmenite lattice is as thin as 1.0 nm, which is smaller than typical $c$-axis parameters of ilmenite-type oxides (approximately 1.4 nm). This sub-unit-cell thick Mn–Ir–O layer admits two $IrO_6$ planes at most, indicating the fragile crystalline phase of Mn–Ir–O. Sandwiching between stable $MnTiO_3$ layers in the superlattice structure stabilizes the ilmenite-type atomic ordering of Mn–Ir–O ultrathin layers.

**Microstructural characterization by scanning transmission electron microscopy (STEM).**
Our concept to form honeycomb-lattice $IrO_6$ planes was further evidenced by cross-sectional STEM observation (Methods). In a [$MnTiO_3$/Mn−Ir−O]$_8$/$MnTiO_3$(buffer) film (no Mn−Ir−O cap, see Fig. S2 for the XRD pattern), eight bright layers running parallel to the film plane are detected in the wide-area high-angle annular dark-field (HAADF) image, shown in Fig. 3(a).

Owing to the Z-contrast nature (Z: atomic number)[23], the bright layers must contain much Ir that has the largest Z among the constituent elements (Mn, Ti, Ir, and O). Also, there are tiny bright island-like regions possibly because of segregated Ir-rich impurities. A close inspection on the single unit of the superlattice (Fig. 3(b)) reveals that a Mn−Ir−O layer between $MnTiO_3$ layers contains a couple of bright atomic planes along the [0001] direction. Using the atomically resolved HAADF image of Fig. 3(c) (the area marked by yellow dashed lines in Fig. 3(b)), we compared each atomic site with a model structure (Fig. 3(d)). It was reported for $MnTiO_3$ that the $MnO_6$ ($TiO_6$) plane has a larger (smaller) atomic displacement along the [0001] direction between intra-plane Mn (Ti) ions[24]. By measuring the atomic displacement in the image, we identified that Ir atoms are located at the *B*-site, which is occupied with Ti atoms in $MnTiO_3$, as schematically shown in Fig. 3(d) (also see the left of Fig. 3(c)). In addition, the bright Ir-containing planes are regularly sandwiched between $MnO_6$ planes, with having similar dumbbell-like characteristic atomic arrangement. On the basis of these results, we conclude that a honeycomb lattice of edge-sharing $IrO_6$ octahedra crystallizes in the Mn−Ir−O layer with ilmenite-type atomic ordering. Note here that the Mn−Ir−O layer composed of two pairs of $IrO_6$ and $MnO_6$ planes corresponds to the 2/3 unit cell (u.c.) of the ilmenite lattice (Fig. 1(a)). We also conducted electron energy loss spectroscopy (EELS) for Mn *L*, Ti *L*, and O *K* edges in the $MnTiO_3$ buffer and Mn−Ir−O layer regions (Fig. S3). While the Mn *L*-edge spectra agree well between the two regions (i.e., $Mn^{2+}$), partial reduction of $Ti^{4+}$ to $Ti^{3+}$ occurs around the Mn−Ir−O layer[25,26]. Associated with this reduction, the existence of oxygen deficiency is suggested from the O *K*-edge spectra. We infer that $Mn^{2+}$ and $Ir^{4+}$ are stable under the oxygen pressure of 10 mTorr used for the deposition, whereas $Ti^{4+}$ is slightly reduced to $Ti^{3+}$. Charge transfer between *B*-site $TiO_6$ and $IrO_6$ planes via *A*-site $MnO_6$ planes may be relevant to this point, though it is not clear at this stage.

**Investigation of surface magnetic order by spin Hall magnetoresistance measurements.**
Judging from these structural analyses, a honeycomb lattice of edge-sharing $IrO_6$ octahedra should partly form a two-dimensional network in the superlattice. While some iridates are known to exhibit metallic conduction[27–31], our superlattice samples were highly insulating (not shown), suggesting the non-metallic electronic structure of Mn−Ir−O as well as the antiferromagnetic insulator $MnTiO_3$. Moreover, with experimental methods applied to bulk crystals[10,13–15], it is difficult to evaluate magnetism in ultrathin films, particularly for cases of antiferromagnetism and quantum spin liquids. Thus, we studied the surface magnetic order using the spin Hall magnetoresistance (SMR) method, which enables electrical characterization of ferromagnetic[32] and antiferromagnetic transitions[33–35], and magnetic anisotropy. By measuring SMR arising at the interface of Pt and $MnTiO_3$ ultrathin films, we have recently demonstrated that the bulk Néel temperature $T_N$ (~ 63 K)[36] and the uniaxial magnetic anisotropy along the *c*-axis direction persist down to the film thickness of 2.9 nm ~ 2 u.c. with six $MnO_6$ planes (thin-$MnTiO_3$ monolayer in Fig. S4)[22]. For the present experiment, we examined five different heterostructures: Pt/4.3-nm-thick MnTiO$_3$ (thick-MnTiO$_3$ monolayer), Pt/1.0-nm-thick Mn−Ir−O/4.3-nm-thick MnTiO$_3$ (bilayer), and Pt/1.0-nm-, 1.9-nm-, and 4.3-nm-thick MnTiO$_3$/1.0-nm-thick Mn−Ir−O/4.3-nm-thick MnTiO$_3$ samples (trilayer A, B, and C, respectively). Schematic structures and XRD patterns of the samples are given in Figs. S4 and S5, respectively. As shown in the inset of Fig. 4(a) (Methods), resistances of the two orthogonally arranged Pt channels ($R_1$ for channel 1 and $R_2$ for channel 2) were simultaneously measured under an in-plane magnetic field $H$ applied parallel to channel 1. Figures 4(a), (b), and (c) show the temperature ($T$) dependence of the field-induced resistance variation, $\Delta R = (R_1 / R_2)$ ($\mu_0 H = 0.5$ T) − ($R_1 / R_2$) ($\mu_0 H = 0$ T), where $\mu_0$ is vacuum permeability. When antiferromagnetic transition takes place in the layer beneath the Pt layer, distinct SMR responses of $R_1$ and $R_2$ below and above $T_N$ result in a kink of $\Delta R$ (refs. [22,33,34]). In fact, $\Delta R$ for both the thick-MnTiO$_3$ monolayer (Fig. 4(a)) and the trilayer C (purple in Fig. 4(c)) exhibits the sudden increase around $T = 60$ K upon heating, around which bulk MnTiO$_3$ undergoes the antiferromagnetic transition[36]. These results indicate that antiferromagnetic order develops in these 4.3-nm-thick MnTiO$_3$ top layers (~ 3 u.c. with nine MnO$_6$ planes) regardless of the underlying Al$_2$O$_3$(0001) substrate or the 1.0-nm-thick Mn−Ir−O/4.3-nm-thick MnTiO$_3$/Al$_2$O$_3$(0001) structure. In stark contrast, $\Delta R$ for the bilayer (Fig. 4(b)), and the trilayer A and B (green and light blue in Fig. 4(c)) exhibits the monotonous increase with increasing $T$ without clear anomalies. Although $A$-site spinful MnO$_6$ planes are common to all samples, magnetic ordering behavior is thus completely different, which we believe reflects magnetic interactions in Mn−Ir−O, i.e., the unique stacking of honeycomb-lattice $IrO_6$ and $MnO_6$ planes.

**Discussion and Conclusions**

Comparing the SMR results, we discuss possible magnetic properties of Mn−Ir−O. In the thick-$MnTiO_3$ monolayer and the trilayer C terminated with 3-u.c.-thick $MnTiO_3$ top layers, nine $MnO_6$ lie beneath the Pt layer. The clearly detected $T_N$ (the kink in $\Delta R$) in these thick samples is consistent with the robust antiferromagnetic ordering in the thin-$MnTiO_3$ monolayer that we reported previously[22]. The absence of such signatures in the bilayer point to local modification of super-exchange interactions in $MnO_6$ planes by intervening $IrO_6$ planes. Taking into account strong spin-orbit coupling inherent to heavy Ir (refs. [1,27–31,37–39]), the disturbance of the antiferromagnetic order is conceivable. In Fig. 4(d), a tentative picture for spin interactions between the Mn sites (black arrows) in the bilayer is shown schematically, where the antiferromagnetic order develops in the $MnTiO_3$ bottom layer but not in the overgrown Mn−Ir−O layer. In the thin $MnTiO_3$ top layers (<< 2 u.c.) of the trilayer A and B, a similar spin fluctuation is probably caused by the neighboring $IrO_6$ planes.

Extrinsic contributions like the surface roughness and the size effect (refs. [40–42]) to the suppression of antiferromagnetic order in the thin $MnTiO_3$ (<< 2 u.c.) and Mn−Ir−O top layers should also be considered. Firstly, all samples measured have smooth surfaces with root-mean-square roughness values of 0.2–1.0 nm, which cannot not solely account for the distinct SMR responses. Another support to the negligible role of surface roughness is that the 2-u.c.-thick thin-MnTiO3 monolayer, despite its island-like film morphology with increased surface roughness, exhibits the bulk-like $T_N$ as in thicker and much flatter samples ($\geq$ 3 u.c.)[22]. Secondly, the total film thicknesses of the heterostructures (including MnTiO3 and Mn−Ir−O layers) are much thicker than the 2 u.c. so that a sufficiently large volume of the Mn sublattice is ensured for the whole heterostructure. This also helps minimize the possible influence of domain disconnection, which generally becomes pronounced in ultrathin films. In stark contrast, the systematic recovery of antiferromagnetic SMR responses with an increase of the thickness of MnTiO3 top layer, from the trilayer A (2/3 u.c.), B (4/3 u.c.), to C (3 u.c.), signals the intrinsic origin that is responsible for spin interactions in a characteristic length. Notably, the striking difference between the antiferromagnetic 2-u.c.-thick thin-MnTiO3 monolayer[22] and the non-antiferromagnetic trilayer B (4/3-u.c.-thick MnTiO3 and 2/3-u.c.-thick Mn−Ir−O (2 u.c. in total) on the MnTiO3 buffer) manifests that the insertion of honeycomb-lattice IrO6 planes dramatically affect spin interactions between the nearby Mn sites. Such a spin-disordered (or spin-frustrated) state is possibly linked to quantum spin liquids, though the experimental identification of the featureless magnetic ground state in ultrathin films poses a great challenge[43–45]. Understanding the spin structures (including the Ir sites) and the excited-state properties in Mn−Ir−O with recently advanced diagnostics like Raman spectroscopy[1,5,46] and in-plane spin transport measurements[47–50] will be the next step. Nevertheless, the suppression of the strong antiferromagnetic order in the Mn sublattice is promising in showing the feasibility of control of spin interactions by artificially engineering ilmenite-type oxides.

In summary, we have materialized a honeycomb lattice of edge-sharing $IrO_6$ octahedra by superlattice formation with ilmenite-type $MnTiO_3$. The systematic SMR measurements suggested the absence of antiferromagnetic order in the surface Mn−Ir−O layer grown on antiferromagnetic $MnTiO_3$. Even though $A$-site $MnO_6$ planes are common to these oxides, their spin interactions between the Mn sites are quite different. A spin fluctuation induced by strong spin-orbit interactions in $IrO_6$ planes may disturb the antiferromagnetic ordering in the Mn−Ir−O layer and the neighboring regions of the $MnTiO_3$ layers. The stabilization of a two-dimensional $IrO_6$ honeycomb lattice by superlattice technique, as well as potential control of the magnetism via dimensionality and the proximity effect, will trigger the development of ilmenite-based Kitaev materials producing exotic physical phenomena.

**Methods**

**Thin-film growth.** A Mn−Ir−O target was prepared from $MnO_2$ and $IrO_2$ powders by spark-plasma-sintering at 50 MPa and 900 °C. The target composition as measured by EDX was Mn : Ir = 0.90 : 1 (atomic ratio). A $MnTiO_3$ buffer layer was grown at a substrate temperature of 850 °C and an oxygen pressure of 10 mTorr using a Mn−Ti−O target[22]. Subsequently, Mn−Ir−O and $MnTiO_3$ layers were alternately deposited for $n$ cycles at 800 °C and 10 mTorr for the formation of a [Mn−Ir−O/$MnTiO_3$]$_n$ superlattice. The crystal structure and composition of the films were characterized by XRD using Cu $K_\alpha$ radiation and EDX, respectively.

**SMR measurements.** A Pt film with a thickness of approximately 2 nm was deposited on the film surface by radio-frequency magnetron sputtering at 150 °C. The heterostructured film was then patterned into an L-shaped multi-terminal device structure using photolithography and Ar-ion milling, followed by electron-beam evaporation of Au/Ti electrodes. Resistance was measured by the four-probe method using a semiconductor parameter analyzer (Agilent 4155C) and nano-volt meters (Keithley 2182A) in a physical property measurement system (Quantum Design Inc.) equipped with a one-axis sample rotator. Details for the measurement scheme and analysis were reported in ref. [22].

**Electron microscopy.** To obtain an electron-transparent thin specimen, the grown thin film with the substrate was mechanically polished, and Ar-ion beam milling was performed at 0.5 kV in the final stage. For the atomic and electronic structure analyses, an aberration corrected STEM system (JEOL ARM300CF) was used, equipped with a DELTA corrector, a cold field emission gun, and an EELS spectrometer (Quantum, Gatan Inc.), operated at 300 kV. The probe forming aperture was 30 mrad and the collection semi-angle for HAADF was 85–200 mrad.

**References**

1. Takagi, H., Takayama, T., Jackeli, G., Khaliullin, G. & Nagler, S. E. Concept and realization of Kitaev quantum spin liquids. *Nat. Phys. Rev.* **1**, 264–280 (2019).

2. Kitaev, A. Anyons in an exactly solved model and beyond. *Ann. Phys.* **321**, 2–111 (2006).

3. Jackeli, G. & Khaliullin, G. Mott Insulators in the Strong Spin-Orbit Coupling Limit: From Heisenberg to a Quantum Compass and Kitaev Models. *Phys. Rev. Lett.* **102**, 017205 (2009).

4. Savary, L. & Balents, L. Quantum spin liquids: a review. *Rep. Prog. Phys.* **80**, 016502 (2017).

5. Motome, Y. & Nasu, J. Hunting Majorana Fermions in Kitaev Magnets. *J. Phys. Soc. Jpn.* **89**, 012002 (2020).

6. Sears, J. A. *et al*. Magnetic order in $\alpha$-RuCl$_3$: A honeycomb-lattice quantum magnet with strong spin-orbit coupling. *Phys. Rev. B* **91**, 144420 (2015).

7. Cao, H. B. *et al*. Low-temperature crystal and magnetic structure of $\alpha$-RuCl$_3$. *Phys. Rev. B* **93**, 134423 (2016).

8. Singh, Y. & Gegenwart, P. Antiferromagnetic Mott insulating state in single crystals of the honeycomb lattice material Na$_2$IrO$_3$. *Phys. Rev. B* **82**, 064412 (2010).

9. Singh, Y. *et al.* Relevance of the Heisenberg-Kitaev Model for the Honeycomb Lattice Iridates $A_2$IrO$_3$. *Phys. Rev. Lett.* **108**, 127203 (2012).

10. Biffin, A. *et al*. Unconventional magnetic order on the hyperhoneycomb Kitaev lattice in

$\beta$-Li$_2$IrO$_3$: Full solution via magnetic resonant x-ray diffraction. *Phys. Rev. B* **90**, 205116 (2014).

11. Takayama, T. *et al*. Hyperhoneycomb Iridate $\beta$-Li$_2$IrO$_3$ as a Platform for Kitaev Magnetism. *Phys. Rev. Lett.* **114**, 077202 (2015).

12. Modic, K. A. *et al*. Realization of a three-dimensional spin–anisotropic harmonic honeycomb iridate. *Nat. Commun.* **5**, 4203 (2014).

13. Biffin, A. *et al*. Noncoplanar and Counterrotating Incommensurate Magnetic Order Stabilized by Kitaev Interactions in $\gamma$-Li$_2$IrO$_3$. *Phys. Rev. Lett.* **113**, 197201 (2014).

14. Kitagawa, K. *et al*. A spin–orbital-entangled quantum liquid on a honeycomb lattice. *Nature* **554**, 341–345 (2018).

15. Kasahara, Y. *et al*. Majorana quantization and half-integer thermal quantum Hall effect in a Kitaev spin liquid. *Nature* **559**, 227–231 (2018).

16. Schlom, D. G., Chen, L.-Q., Pan, X., Schmehl, A. & Zurbuchen, M. A. A Thin Film Approach to Engineering Functionality into Oxides. *J. Am. Ceram. Soc.* **91**, 2429–2454 (2008).

17. Bhattacharya, A. & May, S. J., Magnetic Oxide Heterostructures. *Annu. Rev. Mater. Res.* **44**, 65–90 (2014).

18. Boschker, H. & Mannhart, J. Quantum-Matter Heterostructures. *Annu. Rev. Condens. Matter Phys.* **8**, 145–164 (2017).

19. Haraguchi, Y. *et al*. Magnetic ordering with an XY-like anisotropy in the honeycomb lattice iridates ZnIrO$_3$ and MgIrO$_3$ synthesized via a metathesis reaction. *Phys. Rev. Materials* **2**, 054411 (2018).

20. Haraguchi, Y. & Aruga Katori, H. Strong antiferromagnetic interaction owing to a large trigonal distortion in the spin-orbit-coupled honeycomb lattice iridate CdIrO$_3$. *Phys. Rev. Materials* **4**, 044401 (2020).

21. Toyosaki, H., Kawasaki, M. & Tokura, Y. Atomically smooth and single crystalline MnTiO$_3$ thin films with a ferrotoroidic structure. *Appl. Phys. Lett.* **93**, 072507 (2008).

22. Miura, K., Fujiwara, K., Shiogai, J., Nojima, T. & Tsukazaki, A. Electrical detection of the antiferromagnetic transition in MnTiO$_3$ ultrathin films by spin Hall magnetoresistance. *J. Appl. Phys.* **127**, 103903 (2020).

23. Kim, E. S. & Jeon, C. J. Microwave dielectric properties of ATiO3 (A = Ni, Mg, Co, Mn) ceramics. *J. Eur. Ceram. Soc.* **30**, 341–346 (2010).

24. S. J. Pennycook, S. J. & Boatner, L. A. Chemically sensitive structure-imaging with a scanning transmission electron microscope. *Nature* **336** 565–567 (1988).

25. Calvert, C. C., Rainforth, W. M., Sinclair, D. C. & West, A. R. Characterisation of Grain Boundaries in CaCu$_3$Ti$_4$O$_{12}$ using HREM, EDS and EELS. *J. Phys: Conf. Ser.* **26**, 65–68 (2006).

26. Moatti, A., Sachan, R., Gupta, S. & Narayan, J. Vacancy-Driven Robust Metallicity of

Structurally Pinned Monoclinic Epitaxial VO$_2$ Thin Films. *ACS Appl. Mater. Interfaces* **11**, 3547–3554 (2019).

27. Matsuhira, K. *et al*. Metal–Insulator Transition in Pyrochlore Iridates *Ln*$_2$Ir$_2$O$_7$ (*Ln* = Nd, Sm, and Eu). *J. Phys. Soc. Jpn.* **76**, 043706 (2007).

28. Cao, G. *et al*. Non-Fermi-liquid behavior in nearly ferromagnetic SrIrO$_3$ single crystals. *Phys. Rev. B* **76**, 100402(R) (2007).

29. Jang, S. Y. *et al*. The electronic structure of epitaxially stabilized 5d perovskite Ca$_{1-x}$Sr$_x$IrO$_3$ (*x* = 0, 0.5, and 1) thin films: the role of strong spin–orbit coupling. *J. Phys.: Condens. Matter* **22**, 485602 (2010).

30. Takayama, T. *et al*., Spin-orbit coupling induced semi-metallic state in the 1/3 hole-doped hyper-kagome Na$_3$Ir$_3$O$_8$. *Sci. Rep.* **4**, 6818 (2014).

31. Fujioka, J. *et al*. Strong-correlation induced high-mobility electrons in Dirac semimetal of perovskite oxide. *Nat. Commun.* **10**, 362 (2019).

32. Nakayama, H. *et al*. Spin Hall Magnetoresistance Induced by a Nonequilibrium Proximity Effect. *Phys. Rev. Lett.* **110**, 206601 (2013).

33. Schlitz, R. *et al*. Evolution of the spin hall magnetoresistance in Cr$_2$O$_3$/Pt bilayers close to the Néel temperature. *Appl. Phys. Lett.* **112**, 132401 (2018).

34. Iino, T. *et al*. Resistive detection of the Néel temperature of Cr$_2$O$_3$ thin films. *Appl. Phys. Lett.* **114**, 022402 (2019).

35. Lebrun, R. *et al*. Anisotropies and magnetic phase transitions in insulating antiferromagnets determined by a Spin-Hall magnetoresistance probe. *Commun. Phys.* **2**, 50 (2019).

36. Stickler, J. J. Kern, S. Wold, A. & Heller, G. S. Magnetic Resonance and Susceptibility of Several Ilmenite Powders. *Phys. Rev.* **164**, 765 (1967).

37. Kim, B. J. *et al*. Novel $J_{eff}$ = 1/2 Mott State Induced by Relativistic Spin-Orbit Coupling in $Sr_2IrO_4$. *Phys. Rev. Lett.* **101**, 076402 (2008).

38. Hirata, Y. *et al*. Complex orbital state stabilized by strong spin-orbit coupling in a metallic iridium oxide $IrO_2$. *Phys. Rev. B* **87**, 161111(R) (2013).

39. Schaffer, R., Lee, E. K.-H., Yang, B.-J. & Kim, Y. B. Recent progress on correlated electron systems with strong spin–orbit coupling. *Rep. Prog. Phys.* **79**, 094504 (2016).

40. Abarra, E. N., Takano, K., Hellman, F. & Berkowitz, A. E. Thermodynamic Measurements of Magnetic Ordering in Antiferromagnetic Superlattices. *Phys. Rev. Lett.* **77**, 3451–3454 (1996).

41. Park, S. *et al*. Strain control of Morin temperature in epitaxial *α*-$Fe_2O_3$(0001) film. *EPL* **103**, 27007 (2013).

42. Pati, S. P., Al-Mahdawi, M., Ye, S., Shiokawa, Y., Nozaki, T. & Sahashi, M. Finite-size scaling effect on Néel temperature of antiferromagnetic $Cr_2O_3$ (0001) films in exchange-coupled heterostructures. *Phy. Rev. B* **94**, 224417 (2016).

43. Liu, J. *et al*. Heterointerface engineered electronic and magnetic phases of NdNiO$_3$ thin films. *Nat. Commun.* **4**, 2714 (2013).

44. King, P. D. C. *et al*. Atomic-scale control of competing electronic phases in ultrathin LaNiO$_3$. *Nat. Nanotechnol.* **9**, 443–447 (2014).

45. Bovo, L. Rouleau, C. M., Prabhakaran, D., & Bramwell, S.T. Phase transitions in few-monolayer spin ice films. *Nat. Commun.* **10**, 1219 (2019).

46. Wulferding, D., Choi, Y., Lee W., & Choi, K.-Y. Raman spectroscopic diagnostic of quantum spin liquids. *J. Phys.: Condens. Matter* **32**, 043001 (2020).

47. Cornelissen, L. J., Liu, J. Duine, R. A., Youssef, J. B. & vanWees, B. J. Long-distance transport of magnon spin information in a magnetic insulator at room temperature. *Nat. Phys.* **11**, 1022 (2015).

48. Lebrun, R. *et al*. Tunable long-distance spin transport in a crystalline antiferromagnetic iron oxide. *Nature* **561**, 222 (2018).

49. Koga, A., Minakawa, T., Murakami, Y., & Nasu, J. Spin Transport in the Quantum Spin Liquid State in the $S = 1$ Kitaev Model: Role of the Fractionalized Quasiparticles. *J. Phys. Soc. Jpn.* **89**, 033701 (2020).

50. Minakawa, T., Murakami, Y., Koga, A., & Nasu, J. Majorana-mediated spin transport without spin polarization in Kitaev quantum spin liquids. arXiv:1912.10599v1.

51. Momma, K. & Izumi, F. VESTA 3 for three-dimensional visualization of crystal,

volumetric and morphology data. *J. Appl. Cryst.* **44**, 1272–1276 (2011).

**Acknowledgements**

The authors thank K. Harata, S. Ito, and N. Saito for their supports in experiments, Y. Motome for stimulating discussions, and NEOARK Corporation for the use of a maskless lithography system PALET. This work was performed under the Inter-University Cooperative Research Program of the Institute for Materials Research, Tohoku University (Proposal No. 19G0410). This work was supported by JST CREST (JPMJCR18T2) and JSPS KAKENHI (15H058053 and 19H02423). K.N., R.I. and N.S. acknowledge the supports from JSPS KAKENHI (19H05788) and Research Hub for Advanced Nano Characterization (No. 12024046) by MEXT, Japan.

**Author contributions**

K.M. grew the films, and performed the SMR measurements along with K.F. K.N., R.I., and N.S. conducted the STEM analysis. K.F., K.M., and A.T. wrote the manuscript with input from other authors. All authors discussed the results. A.T. supervised the project.

**Competing interests**

The authors declare no competing interests.

**Data Availability**

The data that support the findings of this study are available from the corresponding author upon reasonable request.

**Figure legends**

**Figure 1. Superlattices of ilmenite-type MnTiO$_3$ and Mn−Ir−O. a,** Left: Crystal structure of ilmenite-type Mn$B$O$_3$ viewed along the $[11\bar{2}0]$ direction, drawn by VESTA (ref. [51]). Right: Honeycomb lattice formed by edge-sharing $B$O$_6$ octahedra in $ab$ plane. Black quadrangles represent the unit cell. **b,** Schematic structure of a Mn−Ir−O(cap)/[$d_{MTO}$-nm-thick MnTiO$_3$/$d_{MIO}$-nm-thick Mn−Ir−O]$_n$/MnTiO$_3$(buffer) film grown on Al$_2$O$_3$(0001). **c,** Out-of-plane XRD pattern of a film with $d_{MTO}$ = 4.3 nm (3 u.c.), $d_{MIO}$ = 0.5 nm, and $n$ = 15. The thickness of MnTiO$_3$ buffer layer was 5.7 nm (4 u.c.). Red arrows and asterisks indicate superlattice reflections and forbidden Al$_2$O$_3$(000$\underline{3n}$) reflections, respectively. **d,** Reciprocal space mapping around Al$_2$O$_3$(02$\bar{2}$$\underline{10}$). (e) In-plane $\phi$ scan results for $(10\bar{1}4)$ peaks of Al$_2$O$_3$ substrate (top) and the film (bottom).

**Figure 2. Examination of the critical $d_{MIO}$ value. a,b,c,** Out-of-plane XRD patterns for $d_{MIO}$-nm-thick Mn−Ir−O(cap)/[4.3-nm-thick MnTiO$_3$/$d_{MIO}$-nm-thick Mn−Ir−O]$_{15}$/5.7-nm-thick MnTiO$_3$(buffer) films with $d_{MIO}$ values of (a) 0.5 nm, (b) 1.0 nm, and (c) 1.4 nm.

**Figure 3. Scanning transmission electron microscopy analysis.** HAADF-STEM images of a [MnTiO$_3$/Mn−Ir−O]$_8$/MnTiO$_3$(buffer) film (without a Mn−Ir−O cap layer) viewed along the Al$_2$O$_3[11\bar{2}0]$ direction: **a,** wide area, **b,** the MnTiO$_3$/Mn−Ir−O/MnTiO$_3$ interface region, and **c,** the atomically resolved images (the area marked by yellow dashed lines in (b)). **d,** Schematic crystal structure corresponding to (c).

**Figure 4. Spin Hall magnetoresistance measurements. a,b,c,** Field-induced variation ($\mu_0 H$ = 0.5 T) in the resistance ratio of two orthogonally arranged Pt channels, $R_1 / R_2$, for (a) 4.3-nm-thick MnTiO$_3$ monolayer (thick-MnTiO$_3$), (b) 1.0-nm-thick Mn−Ir−O/4.3-nm-thick MnTiO$_3$ bilayer (bilayer), and (c) 1.0-nm-, 1.9-nm-, and 4.3-nm-thick MnTiO$_3$/1.0-nm-thick Mn−Ir−O/4.3-nm-thick MnTiO$_3$ trilayer samples (trilayer A, B, and C, respectively). Inset in (a) shows the measurement setup, where the $R_1$ and $R_2$ are measured simultaneously under an in-plane magnetic field $H$ applied parallel to channel 1 with $R_1$ (perpendicular to channel 2 with $R_2$). $I_+$ and $I_-$ are the electrodes used for current injection. **d**, Schematic for the spin ordering at Mn sites in the Mn−Ir−O/MnTiO$_3$ bilayer sample expected from the SMR responses. Upper panel: metal-oxygen octahedra in *ab* plane. Small red spheres represent O ions. Yellow colored octahedra are located above gray colored octahedra. Lower panel: Cross-sectional view along the $[11\bar{2}0]$ direction. Pink, light blue, and brown spheres represent Ir, Mn, and Ti ions accommodated in the oxygen octahedra, respectively. Only localized moments of Mn ions are depicted by black arrows.

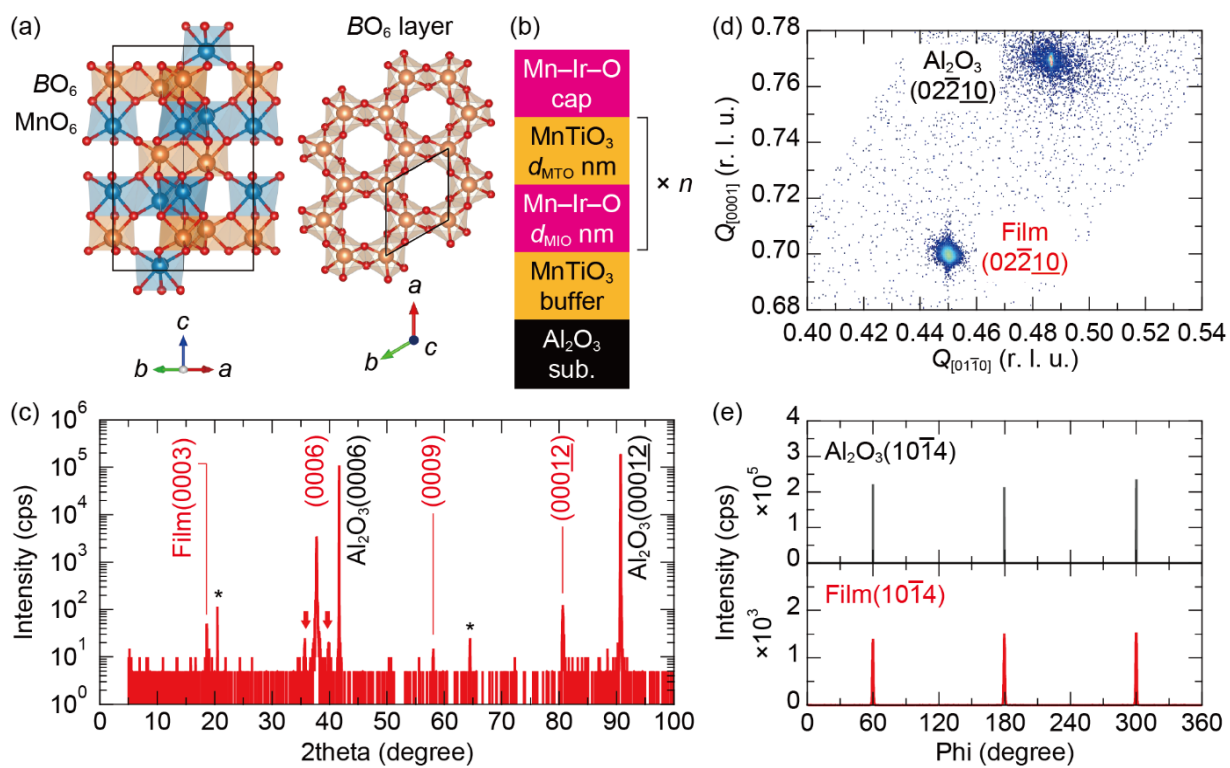

Figure 1

K. Miura *et al.*

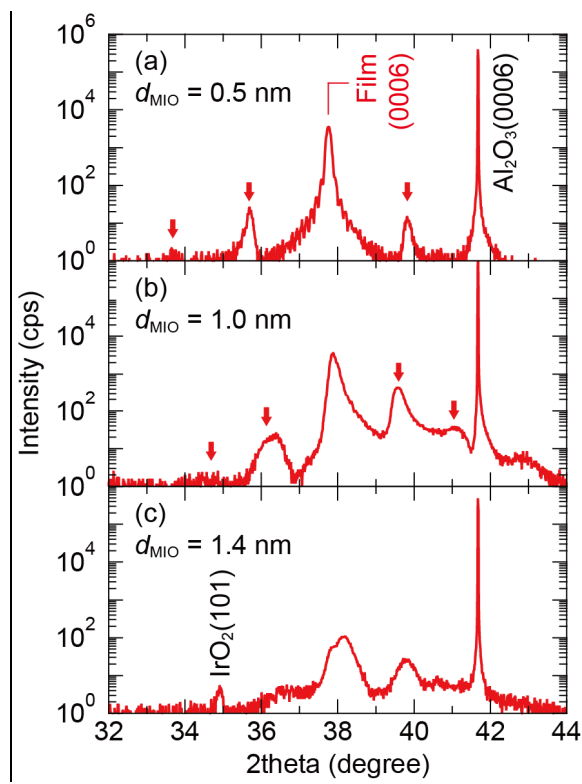

Figure 2

K. Miura *et al.*

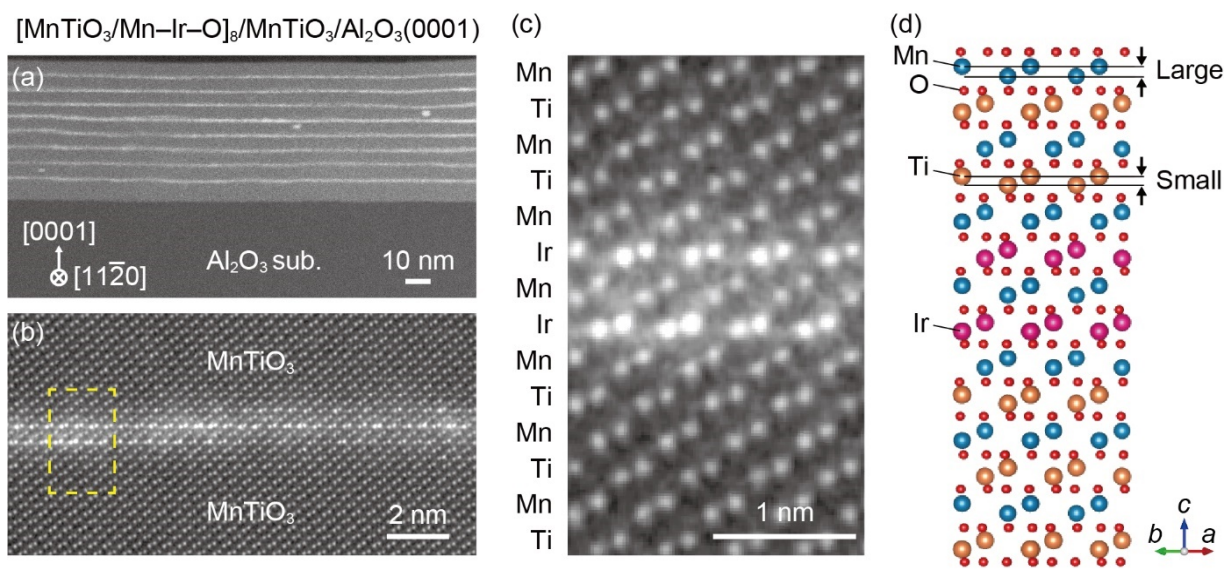

Figure 3

K. Miura *et al.*

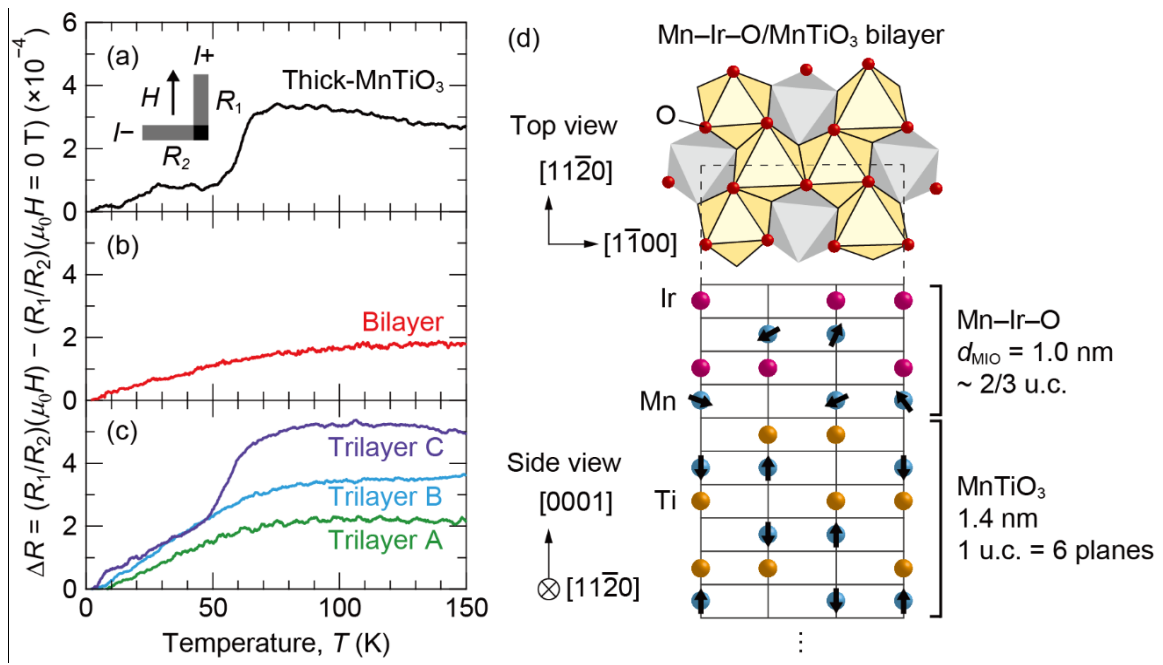

Figure 4

K. Miura *et al.*